\documentclass[a4paper,9pt]{article}
\usepackage[utf8]{inputenc}
\usepackage[T1]{fontenc}
\usepackage{lmodern,textcomp}
\usepackage[english]{babel}
\usepackage[top=2.5cm,bottom=2.5cm,left=2.5cm,right=2.5cm,headsep=1cm]{geometry}%definit la marge
\usepackage{amsfonts,amsmath,amssymb}
\usepackage{color}
\usepackage{graphicx}%pour inclure des images
\usepackage{caption}
\usepackage{subfigure}
\usepackage{array}
\usepackage{multirow}
\usepackage{pict2e}
\usepackage{fancyhdr}%pour parametrer en tete et pied de page
\title{Real time observation of granular rock analogue material deformation and failure using nonlinear laser interferometry}

\author{Pierre Walczak$^{1,*}$, Francesco Mezzapesa$^{1}$, Abderrahmane Bouakline$^{2}$, Julien Ambre$^{2}$,\\Stéphane Bouissou$^{2}$,Stéphane Barland$^{1}$}

\date{%
\small{$^{1}$Universit\'e Côte d'Azur, CNRS, Institut de Physique de Nice, Sophia Antipolis, France\\%
$^{2}$Géoazur, Université de Nice-Sophia Antipolis, France}\\%
$^{*}$Corresponding author: pierre.walczak@inphyni.cnrs.fr\\%
\today}

\begin{document}

\maketitle
\thispagestyle{fancy}

\textbf{A better understanding and anticipation of natural processes such as landsliding or seismic fault activity requires detailed theoretical and experimental analysis of rock mechanics and geomaterial dynamics. These last decades, considerable progress has been made towards understanding deformation and fracture process in laboratory experiment on granular rock materials, as the well-known shear banding experiment. One of the reasons for this progress is the continuous improvement in the instrumental techniques of observation. But the lack of real time methods does not allow the detection of indicators of the upcoming fracture process and thus to anticipate the phenomenon. Here, we have performed uniaxial compression experiments to analyse the response of a granular rock material sample to different perturbations or shocks. We design and use a novel interferometric laser sensor based on the nonlinear self-mixing interferometry technique to observe in real time the deformations of the sample and assess its usefulness as a diagnostic tool for the analysis of geomaterial dynamics. Due to the high spatial and temporal resolution of this approach, we observe both vibrations processes in response to a dynamic loading and the onset of failure. The latter is preceded by a continuous variation of vibration period of the material. After several shocks, the material response is no longer reversible and we detect a progressive accumulation of irreversible deformation leading to the fracture process. We demonstrate that material failure is anticipated by the critical slowing down of the surface vibrational motion, which may therefore be envisioned as an early warning signal or predictor to the macroscopic failure of the sample or of any structure or rock cliff. The nonlinear self-mixing interferometry technique provides formidable tool to observe in real time the direction and amplitude of the motion of the sample surface at microsecond time scales. It is compact, robust and readily extensible to fault propagation measurements. As such, it opens a new window of observation for the study of geomaterial deformation and failure.}

\section*{Key Points}
\begin{itemize}
\item Granular rock material failure is analysed by a new sensing method
\item Out-of-plane displacements are measured thanks to laser self-mixing interferometry
\item Real time geomaterial deformations before and during failure are detected
\item Failure is anticipated by critical slowing down of the sample’s response to shocks
\end{itemize}

\section*{Keywords}
Fracture; rocks physics; irreversible deformation; nonlinear interferometry; granular sample

\section*{Abbrevations}
NLI, Nonlinear Laser Interferometry

\section{Introduction}

Detailed analysis of inelastic strain of rock under stress is a subject of widespread interest. In particular, it is of first importance to build up adequate theory of inelastic deformation and fracturing/failure of geomaterials. The latter is crucial to have a better understanding of natural processes such as landsliding or seismic fault activity. It is also fundamental to predict rock mass response linked to anthropic activities. To achieve these goals, inelastic strain initiation and evolution has been widely observed in stressed rock in the laboratory by several methods. Among these, X-ray tomography\cite{lenoir:2007,besuelle:2010}, monitoring of acoustic emission caused by microcracking activity\cite{lockner:1991} or digital image correlation technique\cite{viggiani:2008,dautriat:2011,nguyen:2011} are the more advanced. In spite of their interest, these techniques do not allow to access the 3-D strain tensor at any point of the sample during the loading process. In order to increase the resolution of observation techniques, optical method have been used as speckle or Moiré Interferometry\cite{grediac:2004}. Recently, in the context of biaxial compression of a granular material, Le Bouil et al.  have used a new optical technique with a high resolution to study the precursors of failure\cite{leBouil:2014}. But these techniques used in the context of failure and deformation observation do not operate in real time.

In this paper, we have used a nonlinear laser interferometry (NLI) method to observe, in real time, discrete out-of-plane displacement with very high accuracy and bandwidth at some points of the sample surface. This technique is here applied to a rock analogue sample subjected to uniaxial compression test where axial shocks are superimposed. Thanks to the NLI method, we have observed that macroscopic failure of the sample is preceded by both period vibration of the material and irreversible deformation. First, we present the principle of self-mixing interferometry, specially used to observe in real time the deformation. Then, we discuss the experimental protocol, and we discuss the results of a typical experiment, showing the advantage of this method.

\section{Methods}
\subsection{Self-mixing interferometry}
\begin{figure}[htb]
\centering
\includegraphics[scale=0.65]{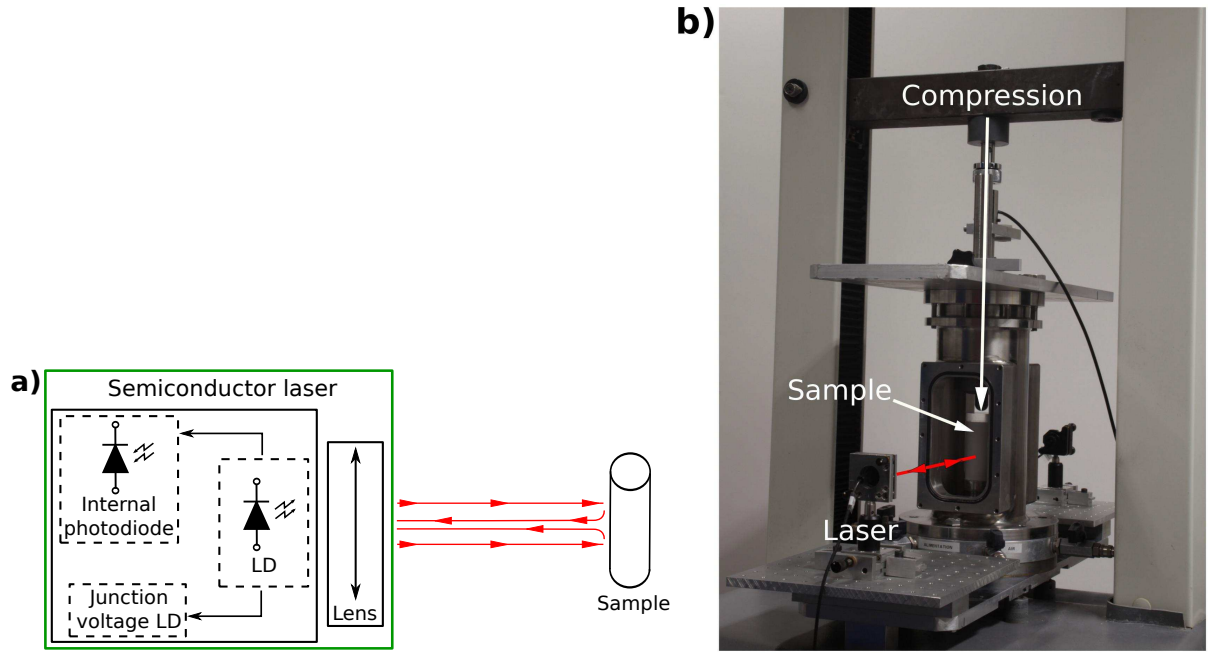}
\caption{Experimental setup of self-mixing interferometry. a) Schematic representation of self-mixing interferometry using a semiconductor laser. LD: Laser Diode. b) Experimental arrangement.}
\label{fig1:setup}
\end{figure}

The Self-mixing interferometry is an instrumentation scheme whose first demonstration has been probably made in 1968\cite{rudd:1968}. In a conventional Michelson interferometer, a coherent light beam emitted by a laser is collimated by a lens, divided and pointed on a target and on a reference mirror. The interferometric pattern composed by the superposition between the light reflected on the target and on the reference mirror is collected on a photodiode. In a self-mixing scheme, it is possible to use the sensitivity of a semiconductor laser to feedback in order to create an interferometric sensor, as we can see in the Fig.\ref{fig1:setup}a. In this case, the laser is used both as a source and a detector where the interference take place. The self-mixing signal in the laser diode can be measured either directly on an integrated photodiode or by monitoring the variations of the laser diode junction voltage\cite{lim:2006}. This technology is used in a wide variety of applications including among others the measurement of the absolute distance\cite{beheim:1986}, the velocity\cite{giuliani:2002} or the displacement\cite{arriaga:2014} of a target. The compactness, integrated and self-aligned nature of this technology makes it a formidable tool to measure in real time the dynamics of the deformation of rock in various conditions. We applied this technique to measure radial (out-of-plane) displacement during uniaxial compression test on cylindrical samples (Fig.~\ref{fig1:setup}b). We used a hard rock analogue GRAM1 (Granular Rock Analogue Material 1) fabricated from a finely ground powder of TiO$_{2}$ with the average grain size of $~$0.3 $\mu$m. This material has the strength two orders of magnitude smaller than that of the most rocks, but otherwise shows very similar with them inelastic properties\cite{nguyen:2011}.

\subsection{Calibration}
In order to monitor the out-of-plane displacement of the material surface, we built two self-mixing interferometers. Each of them monitoring a very small region of each side of the sample. The monitored area is defined by the laser spot size, which is of the order of a few $\mu m^{2}$. Each interferometer is based on a semiconductor laser diode (Thorlabs ML725B8F) and an adjustable lens mount (Thorlabs LTN330-C). The lasers operate at a wavelength of $\lambda_{0}$ = 1310 nm. They are placed at the same height and horizontal axis on both sides of the sample. We can control the size of the beam thanks to the lens. This allows to adjust the quantity of feedback going back inside the laser. The signal is amplified ($10^{4}$ gain, AC-coupled) and recorded by an oscilloscope. 

In spite of the diffusive nature of the material surface (grain size close to the laser wavelength), enough self-mixing signal is easily obtained and care should be taken to avoid excessive field reinjection since only very weak feedback ($\sim 10^{-4}$) is sufficient for adequate operation (larger reinjection leading easily to complex chaotic regimes\cite{lang:1980,kliese:2014,kane:2005}.

\begin{figure}[htb]
\centering
\includegraphics[scale=0.35]{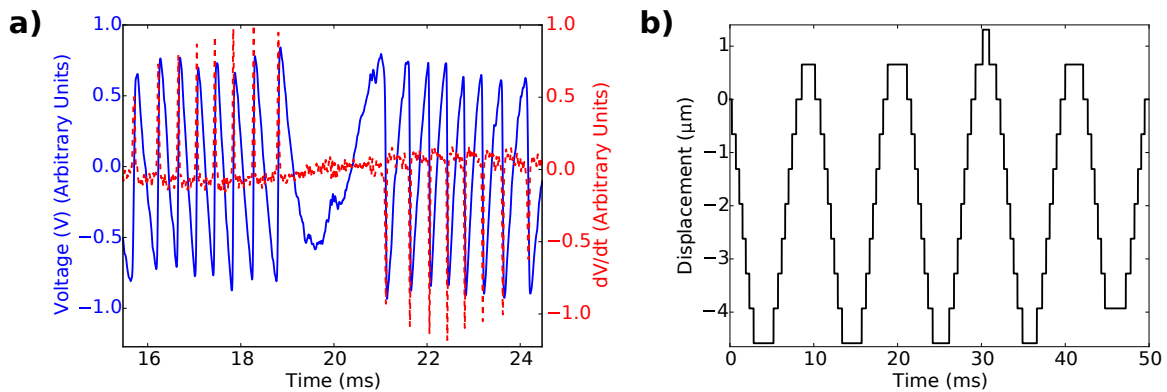}
\caption{(a) Sawtooth signal (blue solid line) and its derivative (red dash line) over one period of sinusoidal relative motion. (b) Displacement reconstruction over five periods of sinusoidal relative motion.}
\label{fig2:calib}
\end{figure}

In the correct operation range, the detected signal is constant when the target is stationary. When the target moves either towards or away from the laser at constant speed, the signal has a sawtooth shape, each upward or downward jump corresponding to a displacement by a half wavelength in one or the other direction. In order to calibrate the signal and check operation range, we have placed the detector on a piezo-actuated translation stage which imposed a sinusoidal relative motion of the sample surface with respect to the detector. The resulting signal is shown in Fig.~\ref{fig2:calib}a (blue solid line). We can see a succession of transitions characterizing the direction of the displacement. Although advanced signal processing techniques may enable very precise trajectory reconstruction well below the laser half-wavelength\cite{arriaga:2014,bes:2006}, we use a simpler, robust and here sufficiently accurate method. The signal is processed by computing numerically its time derivative (dashed red line), which will feature downwards or upwards spikes, which are easily counted via a simple thresholding procedure. The motion is then reconstructed by adding or substracting half-wavelengths depending on the direction of the spike. This is shown on Fig.~\ref{fig2:calib}b, where one easily recognizes the sinusoidal motion imposed by the piezo actuator. After this calibration procedure, the sinusoidal signal imposing periodic motion is switched off.

\section{Surface displacement in response to shocks}
In this section, we describe the observations which were realized following mechanical perturbations (shocks) applied to the sample.
\subsection{Reversible response}
\begin{figure}[htb]
\centering
\includegraphics[scale=0.35]{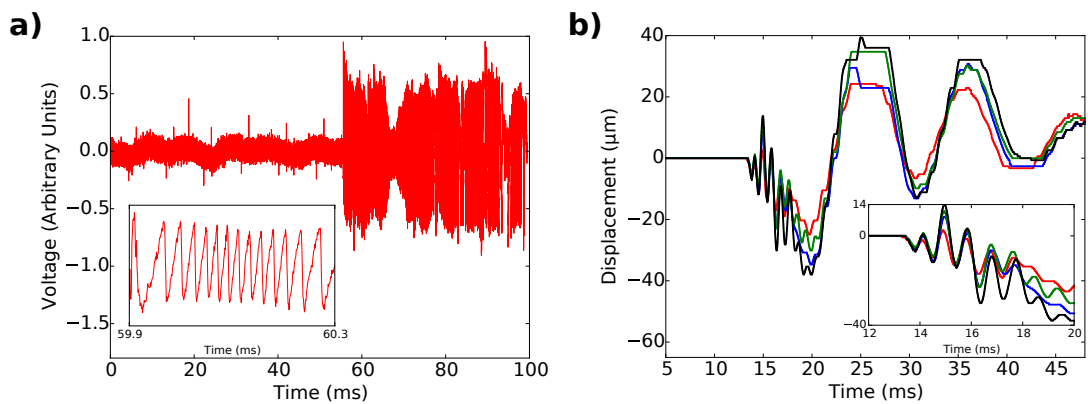}
\caption{We apply different uniaxial perturbations on a rock sample in which a uniaxal compression is applied. a) Signal recorded by the oscilloscope and a zoom in inset of this figure. We can see a sawtooth-like signal after the shock. b) Displacement reconstruction for four different shocks.}
\label{fig3:rir}
\end{figure}

In this experiment, the sample was submitted to a constant force in order to ensure optimal positioning and alignment of the press and the sample. Then several mechanical perturbations were applied by hitting with a hammer the top of the press. Following each of these shocks, we monitored the laser signal and reconstructed the relative displacement of the sample surface with respect to the laser. An example of time trace is shown on Fig.~\ref{fig3:rir}a. Up to about 56 ms the signal is constant indicating absence of relative motion. Then a very complex and fast signal is detected but (see inset) it still essentially consists of abrupt jumps which denote $\lambda /2$ displacement. The amplitude variations result from the variations of the feedback level due to the diffusive surface but the spikes in the derivatives can still easily be identified and the trajectories can be reconstructed. Four realizations of the experiment are displayed on Fig.~\ref{fig3:rir}b. All trajectories are extremely similar and consist of two sets of oscillation, a slow one of about 10 ms period and a much faster one (about 1 ms period). We have performed the same measurement replacing the sample with a comparatively much harder material (wood) and observed that the slow oscillations were still present and identical but the fast ones did not exist. From there we conclude that the slow oscillations are due to the mechanical deformation of the press following the shock but most importantly the fast oscillations correspond to back and forth oscillations of the sample surface, observed in real time. From there we can infer a resonance frequency for compression waves in the sample of about 1 kHz. Although this value is one order of magnitude lower compared to P waves velocity what can be inferred from elastic modulii estimated from static deformation measurements\cite{nguyen:2011}. This difference may be explained by the difference between static and dynamic elastic modulus and also by the damaging of the sample over the successive shocks and/or by the fact that we may be measuring here nonlinear oscillations (which implies dependence of the frequency on the amplitude). In fact, we observe that the damping of these oscillations is not clearly exponential, which suggests that the surface motion is not simply that of a damped linear oscillator and therefore the compression waves we observe (although the material is still in an elastic regime) are not linear.

\subsection{Irreversible response from damaging to failure}
\begin{figure}[htb]
\centering
\includegraphics[scale=0.35]{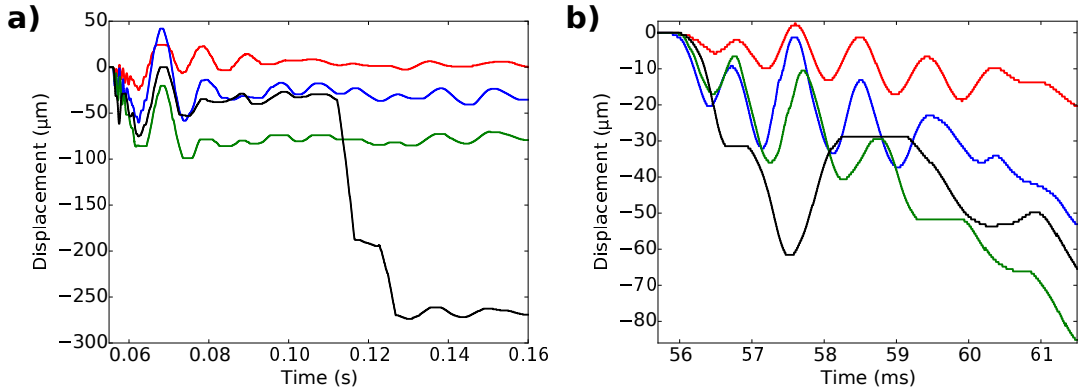}
\caption{Displacement in response to successive perturbations until failure. Red, blue, green and black curves correspond respectively to shocks 1,13,14,15. a) long term evolution, b) zoom on the initial vibrational motion.}
\label{fig4:irr}
\end{figure}

Fig.~\ref{fig4:irr} shows the evolution of time trace recorded for successive shocks. At the beginning of the experiment (1st and 5th shock) no significant difference is observed on the recordings. After the slow oscillation corresponding to the press vibration (Fig.~\ref{fig4:irr}a), the first shocks (in red and blue) have almost the same oscillation period. We can however observe a difference after the damping of the vibrations. The 1st shock does not modify irreversibly the sample and the measured surface returns to its initial position. On the contrary, at the shock number 13 (blue trace), the surface does not return to the initial position, remaining about 30 micrometers away from where it was initially. This indicates already irreversible damage. Furthermore, on Fig.~\ref{fig4:irr}b, detailed analysis of the blue trace may in fact display a slightly delayed fourth and fifth peak as compared to the first response (in red). These observations are even more pronounced on the green trace, shock number 14. On Fig.~\ref{fig4:irr}a, the surface does not return at all to its initial position, remaining close to 100 microns away from it. Correspondingly, Fig.~\ref{fig4:irr}b, the fast initial dynamics shows a markedly different periodicity, the response of the material being noticeably slower. We attribute this to the emergence of microfractures of the sample caused by repetitive shocks. Finally, the sample breaks at shock number 15 (black trace). On Fig.~\ref{fig4:irr}a, the actual breakage takes place with a very long delay (0.11 ms) with respect to the shock. However, this fracture can be anticipated already from the immediate response of the sample, as shown on Fig.~\ref{fig4:irr}b. In this last case (black curve), the motion is very asymmetric and the period is much larger. It is notable that critical slowing down is often proposed as an indicator of upcoming tipping points\cite{lenton:2008,scheffer:2012}. Here the drastically slower response of the sample is followed by the actual failure of the material, which takes place about 50 ms later.

\subsection{Discussion and conclusion}
We have demonstrated that self-mixing interferometry can provide an unexplored viewpoint on the deformations of geomaterials in real time. In addition to being compact and cost-effective, it can easily provide information on surface displacements at the micrometer scale in spite of the diffusive sample surface and it easily covers frequencies up to the MHz range. 

In this work, we have shown the effectivity of this approach in the observation of the response of analogue rock material to shocks, both in the reversible and irreversible regimes. Most importantly, contrary to static measurements, the real time measurement method presented here has allowed the observation of a drastic slowing down of the material response, which can therefore be used as an indicator for imminent failure.

Furthermore, the technique can in principle be used across fluids from the outside of a confining pressure cell if an optical access is available. It can also be used on a rock cliff if the laser diode can be placed at a few centimeters from the target to measure. Further research avenues opened by this approach include the quantitative analysis of amplitude and frequency response to calibrated perturbations and wave or cracks propagation via multipoint measurements.

\section*{Acknowledgments}
We acknowledge support of Université de Nice and Fédération Wolfgang Döblin for financial support through project fPod.

% Bibliography
\bibliographystyle{bibPi}
\bibliography{article}

\end{document}